\documentclass[twocolumn,prl,showpacs,amsmath,amstex,amssymb,mathfonts,superscriptaddress]{revtex4-1}
\pdfoutput=1
\usepackage{amsthm,color,amsfonts,graphicx,verbatim}
\usepackage{amsmath}
\usepackage{amssymb}
\usepackage{amsthm}
\usepackage{amsfonts}
\usepackage{listings}
\usepackage{enumerate}
\usepackage{latexsym}
\usepackage{psfrag}
\usepackage{bm}
\usepackage{graphicx}
\usepackage{hyperref}
\usepackage{soul}
\hypersetup{
    bookmarks=true,         
    unicode=false,          
    pdftoolbar=true,        
    pdfmenubar=true,        
    pdffitwindow=false,     
    pdfstartview={FitH},    
    pdftitle={Local Conservation Laws and the Structure of Many-Body Localized States},    
    pdfauthor={},     
    pdfsubject={},   
    pdfcreator={},   
    pdfproducer={}, 
    pdfkeywords={keyword1} {key2} {key3}, 
    pdfnewwindow=true,      
    colorlinks=true,       
    linkcolor=black,          
    citecolor=blue,        
    filecolor=magenta,      
    urlcolor=blue           
}

\newcommand{\be}{\begin{equation}}
\newcommand{\ee}{\end{equation}}
\newcommand{\bea}{\begin{eqnarray}}
\newcommand{\eea}{\end{eqnarray}}

\newcommand{\la}{\langle}
\newcommand{\ra}{\rangle}

\renewcommand{\phi}{\varphi}
\renewcommand{\epsilon}{\varepsilon}
\renewcommand{\vec}[1]{{\bf #1}}

\begin{document}
\title{Local Conservation Laws and the Structure of the Many-Body Localized States}
\author{Maksym Serbyn} 
\affiliation{Department of Physics, Massachusetts Institute of Technology, Cambridge, MA 02138, USA}
\author{Z. Papi\'c}
\affiliation{Department of Electrical Engineering, Princeton University, Princeton, NJ 08544, USA}
\author{Dmitry A. Abanin}
\affiliation{Perimeter Institute for Theoretical Physics, Waterloo, ON N2L 2Y5, Canada}
\affiliation{Institute for Quantum Computing, Waterloo, ON N2L 3G1, Canada}

\date{\today}
\begin{abstract}

We construct a complete set of local integrals of motion that characterize the many-body localized (MBL) phase.
Our approach relies on the assumption that local perturbations act locally on the eigenstates in the MBL phase, which is  supported by numerical simulations of the random-field XXZ spin chain. We describe the structure of the eigenstates in the MBL phase and discuss the implications of local conservation laws for its non-equilibrium quantum dynamics. We argue that the many-body localization can be used to protect coherence in the system by suppressing relaxation between eigenstates with different local integrals of motion.
\end{abstract}
\pacs{}
\maketitle

{\bf Introduction.} Localization of eigenstates of a single-particle in the presence of disorder is among the most remarkable consequences of quantum mechanics. Although the single-particle localization and localization-delocalization transition are well-understood~\cite{Anderson58, LeeRam}, much less is known about the nature of the eigenstates in {\it interacting} many-body disordered systems. The interest in the problem of the many-body localization was rekindled when recent works ~\cite{Basko06,Mirlin05} suggested the localized phase to be stable with respect to weak interactions. This conjecture was also corroborated by numerical studies~\cite{OganesyanHuse,Znidaric08,Monthus10,Berkelbach10,PalHuse,GogolinMueller,SilvaPRB,RigolPRL,Luca11,Cuevas12,Luca13,we}.  
 
In the non-interacting localized phase dynamics is simple because any initial wave function can be decomposed into a superposition of localized single-particle eigenstates. However, when interactions are introduced, the dynamics becomes notably richer~\cite{Moore12,Znidaric08,we,Altman13}. Although particle transport is still expected to be blocked, the time evolution of initial product states in the interacting localized phase generates a universal slow growth of entanglement entropy~\cite{Moore12}. Saturated entropy was established to be proportional to system size~\cite{Moore12,Znidaric08,we,Altman13}, and such growth of the entanglement was argued to reflect ``partial thermalization" of the system. However, the type of the ensemble describing the MBL phase is unknown. 

On the experimental side, probing the dynamics of interacting disordered systems has become feasible due to the advances in the field of ultracold atomic gases~\cite{BlochColdAtoms,PolkovnikovColdAtoms}. In particular, nearly isolated quantum systems of cold atoms can now be engineered, prepared in a variety of initial states (including product states~\cite{Bloch12}), and studied during their subsequent time evolution. These opportunities call for developing a better understanding of the laws that govern the dynamics in the MBL phase.  
      
Here we consider a many-body system whose eigenstates at all energies are localized, and show that they can be characterized by a large number of emergent {\it local} integrals of motion corresponding to multiple local conservation laws. These integrals of motion form a complete set, in the sense that their values completely determine the eigenstates. 
Local conservation laws strongly constrain the quantum dynamics in the MBL phase, preventing a complete thermalization of any given subsystem. Any initial state can be decomposed in terms of the eigenstates possessing definite values of the integrals of motion. During time evolution, the weights of different states cannot change. However, because of the exponentially weak interaction between distant degrees of freedom, the relative phases between the states with different values of local integrals of motion become randomized. Any local observable at long times is therefore determined only by the set of probabilities of local integrals of motion that affect the degrees of freedom in the region where the observable is measured. We refer to this as the  local diagonal ensemble. The dephasing due to the interactions between distant subsystems is a distinct feature of the MBL phase compared to the non-interacting one, and underlies the slow growth of entanglement~\cite{Chiara06, Igloi12, Moore12, we}.

{\bf Integrals of motion.}  First, we note that for the non-interacting case the local integrals of motion are simply given by $\hat{I}_i=c_i^\dagger c_i$, where $c_i^\dagger$ creates a localized single-particle state. For fermions, the possible eigenvalues of this integral of motion are $I_i \in \{0,1 \}$. In a system with $K$ orbitals, there are $2^K$ eigenstates, which are uniquely labeled by the eigenvalues of $K$ integrals of motion.

In order to explicitly construct local integrals of motion for an interacting system, we assume the following property of the localized phase:
{\it local perturbations lead only to local modifications of the eigenstates in the MBL phase.}
That is, if we act on a MBL eigenstate with a local perturbation, introduced either adiabatically or instantaneously, the degrees of freedom situated at a distance $L\gg \xi$ (here $\xi$ is the localization length~\cite{footnote1}) away from the support of the perturbation operator, are generally affected exponentially weakly. We will support this statement below by the numerical study of the random-field XXZ chain, also considered in Refs.~\cite{Znidaric08,PalHuse,SilvaPRB,Luca13,we}.

Let us consider a MBL system described by a local Hamiltonian, and let us divide it into subsystems of size $l \gg \xi$. Without loss of generality, we consider a 1D system, although our conclusions apply to localized phases in any number of spatial dimensions. We number the subsystems by $i=1,..,N$ from left to right, assuming they are of equal size ${\cal M}$, corresponding to Hilbert space dimension of  each subsystem (e.g., for $K$ spins, ${\cal M}=2^K$). For the fixed subsystem $i$, we denote parts of the full system to the left and to the right of $i$ by ${\cal L}_i$ and ${\cal R}_i$, respectively. The Hamiltonian can be written as: 
\be\label{eq:hamiltonian_i}
H=H_{\cal L}+H_i+H_{\cal R}+H_{{\cal L}i}+H_{{\cal R}i},
\ee
where $H_{\cal L}, H_{\cal R}, H_i$ act only on the degrees of freedom in ${\cal L},{\cal R},i$, while $H_{{\cal L}i}$, $H_{{\cal R}i}$ couple ${\cal L},i$ and ${\cal R},i$.

If the subsystems ${\cal L},i,{\cal R}$ are disconnected from each other (i.e., $H_{{\cal L}i}, H_{{\cal R}i}$ are set to zero), the eigenstates are simple products: 
$|\alpha\beta\gamma\ra_0=|\alpha\ra_{\cal L}\otimes |\beta\ra_i \otimes |\gamma\ra_{\cal R}$, 
where $\alpha\in \{ 1,\ldots,{\cal M}^{i-1} \}$, $\beta=\{1,\ldots,{\cal M} \}$, $\gamma=\{1,\ldots,{\cal M}^{N-i} \}$.
Once the subsystems are connected, the eigenstates of the full Hamiltonian~(\ref{eq:hamiltonian_i}) are obtained from the product states $|\alpha\beta\gamma\ra_0$ by nearly {\it local} rotations. We label the resulting eigenstates by their ``ancestors'', omitting ``0" subscript,  
\be\label{eq:eigenstates_connect}
|\alpha\beta\gamma\ra=\hat O_{{\cal L}i} \hat O_{{\cal R}i} |\alpha\ra_{\cal L}\otimes \beta\ra_i \otimes |\gamma\ra_{\cal R}.  
\ee
Operator $\hat O_{{\cal L}i}$ is a unitary many-body rotation which strongly transforms only the degrees of freedom within a distance $\sim \xi$ away from the boundary between ${\cal L}$ and $i$ (similarly for $\hat O_{{\cal R}i}$). The commutator of $\hat O_{{\cal L}i}$ and $\hat O_{{\cal R}i}$, as well as the action on the degrees of freedom far away, decays exponentially. We note that the assignment (\ref{eq:eigenstates_connect}), which links the eigenstates of the system to the eigenstates of subsystems, is not unique, and assume that a certain one-to-one correspondence is chosen. 

We now define the integral of motion for subsystem $i$:
\be\label{eq:integral}
\hat I_i=\sum_{\beta=1}^{{\cal M}} \beta \sum_{\alpha=1}^{{\cal M}^{i-1}} \sum_{\gamma=1}^{{\cal M}^{N-i}} |\alpha\beta \gamma\ra \la \alpha\beta\gamma|.
\ee
Being a linear combination of projectors onto the exact eigenstates, $\hat I_i$ necessarily commutes with the Hamiltonian and assumes eigenvalues $1,\ldots,{\cal M}$. Intuitively, states with the same eigenvalue of $\hat I_i$ look nearly identical within the subsystem $i$ at distances larger than $\xi$ away from the boundaries with subsystems ${\cal L},{\cal R}$. 

Sums of projectors onto the eigenstates are integrals of motion by construction, however generally such operators are non-local and affect all degrees of freedom of the system. The operator in Eq.~(\ref{eq:integral}) is special in that it is local, i.e., it weakly affects the degrees of freedom in ${\cal L}$ or ${\cal R}$ at a distance $x\gg \xi$ away from the boundaries with the $i$th subsystem. The locality of $\hat I_i$ follows directly from the locality of operators $\hat O_{{\cal L}i}$, $\hat O_{{\cal R}i}$,  which implies that the sum of projectors becomes very close to the identity operator far away from the boundaries. Below, we will test the locality of the operator in Eq.~(\ref{eq:integral}) in a specific model.

Having defined the integral of motion for the subsystem $i$, we can similarly define $N-1$ integrals of motion for the remaining $N-1$ subsystems, such that in total we have $N$ integrals, $\hat I_{i}$, $i=1,\ldots,N$. 
Different $\hat{I}_i$ commute with each other $[\hat I_i, \hat I_j]=0$ since they are sums of projectors onto the exact eigenstates of the full system. Each $\hat I_i$ has $\cal M$ possible eigenvalues, thus the full description of the system via integrals of motion requires $ {\cal M}^N$ parameters, which coincides with the dimensionality of the Hilbert space. An operator $\hat I_i$ can also be viewed as the $z$-component $\hat I_i=\hat S_{iz}$ of a ``spin"  $S=({\cal M}-1)/2$. Raising and lowering operators can then be used to construct the entire set of eigenstates, starting from any given eigenstate $|I_1 I_2 \ldots I_N\ra$ characterized by the integrals of motion $I_1, I_2,\ldots, I_N$. Therefore, specifying the eigenvalues of all integrals of motion defined above completely determines the eigenstates of the system.

{\bf Hamiltonian and its relation to integrals of motion.} The Hamiltonian takes an especially simple form when written in terms of the integrals of motion:
\begin{eqnarray}\label{eq:hamiltonian_integrals}
\nonumber H&=& \sum _{i}^N \sum_{I_1=1}^{\cal M}     E_{I_i}  \hat {\cal P}^i _{I_i} + 
\sum _{i\neq j}^N \sum_{I_i, I_j=1}^{\cal M}     E_{I_i I_j}  \hat {\cal P}^i _{I_i}  \hat {\cal P}^j _{I_j}
\\
&& +\sum _{i<j<k}^N \sum_{I_i, I_j I_k =1}^{\cal M}     E_{I_i I_j I_k}  \hat {\cal P}^i _{I_i}  \hat {\cal P}^j _{I_j}  \hat {\cal P}^k _{I_k}+..., 
\end{eqnarray}
 where $ \hat {\cal P}^i _{I_i}$ is the projector onto the subspace for which the eigenvalue of $i$th integral of motion is equal to $I_i$.
 In the above equation, $E_{I_i}$ can be roughly viewed as the energy of the $i$th subsystem for the sector $I_i$, $E_{I_i I_j}$ is the interaction energy between $i$ and $j$ subsystems, etc. There are interactions between any given $n$ subsystems, however, they are exponentially small. Generally, we expect that energies $E_{I_i}$ are proportional $l$, the size of the subsystems. $E_{I_i I_j}$ are proportional to $\xi$ when $i = j\pm 1$, and are suppressed as $\xi e^{-l(|i-j|-1)/\xi}$  otherwise (the interactions between the neighboring subsystems are limited to the boundary and are therefore proportional to $\xi$). The above representation of the Hamiltonian gives us a way to describe the dynamics in the MBL phase for various kinds of initial states~\cite{GogolinMueller,SilvaPRB,PalHuse,Znidaric08,Moore12,we} 

{\bf Dynamics.} As a first step, we consider the dynamics of an eigenstate which is perturbed locally. 
We assume a sudden action of the local unitary operator $\hat U$ on the eigenstate $|\Psi_0\ra=|I_1 I_2 \ldots I_N\ra$. Operator $\hat U$ acts only on the degrees of freedom in the subsystem 1, and its support is situated far from the boundary between subsystems 1 and 2.  The initial wave function $|\Psi(t=0)\ra$  can be decomposed in terms of the eigenstates:
\be\label{eq:psi_perturbed}
|\Psi(t=0)\ra=\hat{U} |\Psi_0\ra
=
\sum_{I_1'} U_{I_1 I_1'} |I_1' I_2 \ldots I_N\ra + \ldots.
\ee 
This form of the decomposition is dictated by the fact that the values of the integrals of motion $I_2,\ldots,I_N$ can be changed only with an exponentially small probability, hence the terms with other values of $I_2, I_3,\ldots$ in Eq.~(\ref{eq:psi_perturbed}) are represented by ellipses. Neglecting these terms, the subsequent dynamics becomes trivial: 
\be\label{eq:psi_time_evolved}
|\Psi(t)\ra=\sum_{I_1'}U_{I_1 I_1'} e^{-iE_{I_1' I_2 \ldots I_N} t} |I_1' I_2 \ldots I_N\ra,  
\ee
where $E_{I_1' I_2 \ldots I_N}$ is the energy of the state $|I_1' I_2...I_N\ra$. Generally, we expect a finite number of different $I_1'$ which have significant matrix elements $U_{I_1 I_1'}$, typically comparable to the dimensionality of a subsystem of size $\sim \xi$. Therefore, the time evolution (\ref{eq:psi_time_evolved}) describes coherent oscillations that involve  a finite number of states. Any local observable in the region 1 would therefore oscillate at a number of frequencies, showing revivals but no dephasing. This situation changes if the state $|\Psi_0\ra$ is not an eigenstate, but a superposition of several eigenstates which involve different values of $I_2,I_3,...I_k$. In this case, exponentially slow dephasing arises, suppressing the revivals and oscillations of local observables in the long-time limit. The values of observables at long times are determined by the probabilities $|U_{I_1 I_1'}|^2$. 

Second, we describe the global evolution of states which differ from the eigenstates everywhere, not just locally. For definiteness, consider an initial product state of subsystems $1,2,\ldots,N$: 
 \be\label{eq:product}
 |\Psi\ra=\otimes_{i=1}^N \left(\sum_{\alpha_i=1}^{\cal M} A_{\alpha_i} |\alpha_i\ra \right), 
 \ee
 where $|\alpha_i\ra$ is an eigenstate of the Hamiltonian $H_i$. Modern experimental techniques allow for the preparation and manipulation of such states in optical lattices~\cite{Bloch12}. 

Each component $\otimes_{i=1}^N |\alpha_i\ra$ of the product state (\ref{eq:product}) can be related to the eigenstate of the whole system, $|I_1 I_2\ldots I_N\ra$, by the set of local rotations acting near the boundaries between different subsystems. The dynamics corresponding to this effect will be limited to the boundaries between pairs of subsystems. However, for each wave function, degrees of freedom at a distance $x\gg \xi$ away from the boundary will remain undisturbed. Such dynamics therefore does not generate long-range entanglement. 
 
 More importantly, since we are dealing with a superposition of different product states $\otimes_{i=1}^N |\alpha_i\ra$, the degrees of freedom in the subsystem $i$ will be in a superposition of states with different values of the integral of motion $I_i$. Different states entering this superposition are eigenstates, therefore their relative weights cannot change under time evolution. However, their \emph{phases} will become random due to the interactions with distant subsystems, as is evident from the Hamiltonian (\ref{eq:hamiltonian_integrals}).  Such dephasing, though exponentially slow, \emph{will} produce long-range entanglement, and thus give rise to the entanglement entropy that is extensive in the system size and determined by the participation ratios of different eigenstates~\cite{Polkovnikov}, as discussed in detail in Ref.~\onlinecite{we}. 
 

\begin{figure}[htb]
\includegraphics[width=\linewidth]{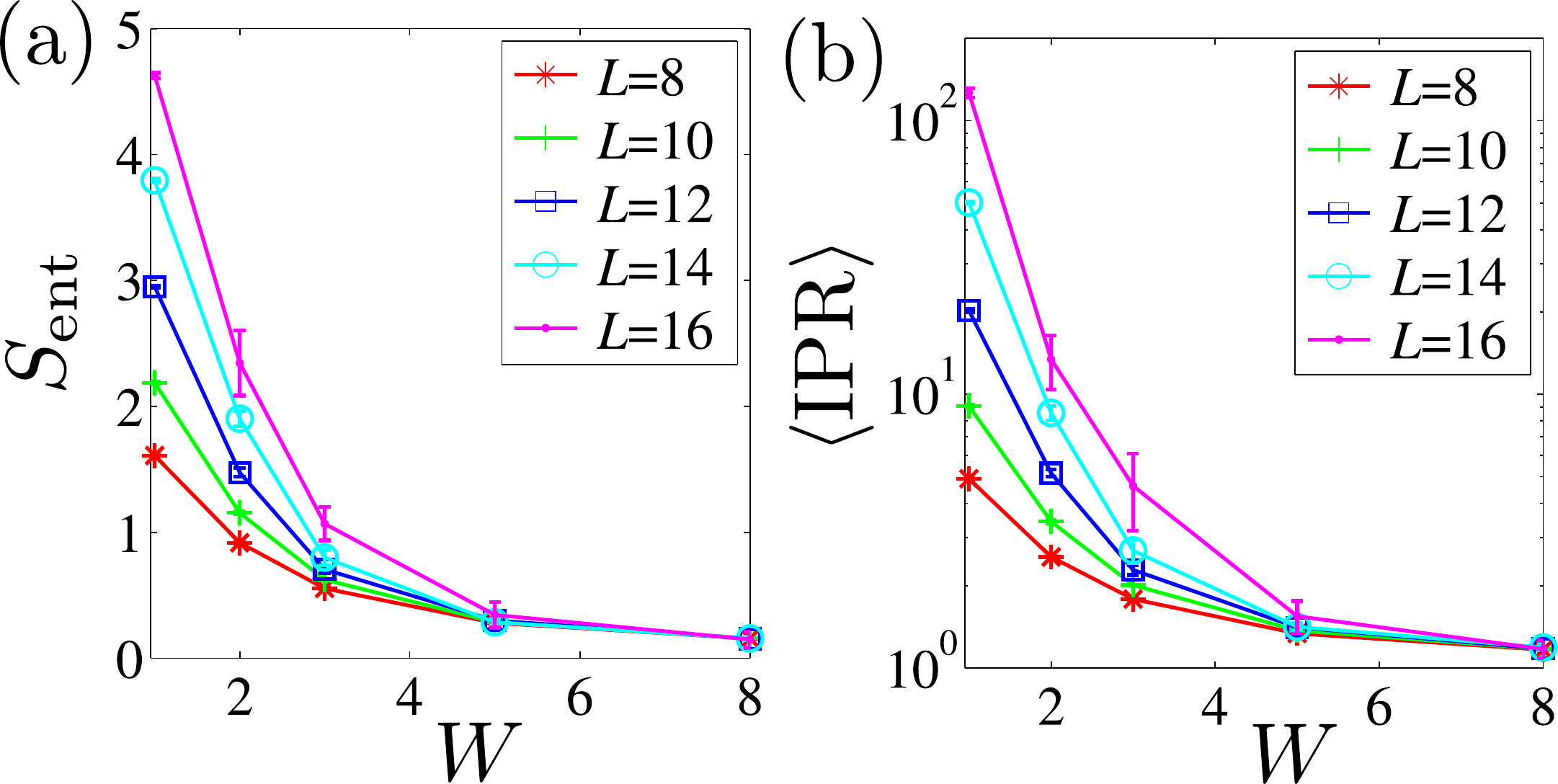}
\caption{ \label{Fig:ipr0}  (a) Averaged entanglement entropy is of the order one and varies weakly with $L$ for strong disorder, indicating that eigenstates are short-range entangled.  Interaction strength is $V = 1$. (b) Inverse participation ratios for the product of two eigenstates of $\mathcal{L}$ and $\mathcal{R}$ subsystems are close to one and do not depend on $L$ for strong disorder.}
\end{figure}
{\bf Numerical simulations.} Although our construction is general, we now test the validity of our basic assumption using exact diagonalization of a particular model -- the random-field XXZ spin chain. We consider a chain of $L$ spins with open boundary conditions, exchange $J_\perp=1$ and interaction strength $J_z=V$, while random on-site magnetic field is uniformly distributed in the interval $ \pm W$. Total $z-$component of the spin is conserved, and calculations are restricted to $S_z=0$ sector. For $V=0$ the model is equivalent to free fermions with disorder, and all states are localized. Because of the limits on the accessible system sizes in exact diagonalization, we restrict ourselves to the case of the symmetric bipartite division of the full system, $\cal LR$, into the left ($\cal L$) and right ($\cal R$) half. 
 
First, we study the averaged entanglement entropy $S_{\rm ent}$ of $\cal L$ subsystem in the eigenstates of $\cal LR$, illustrated in Fig.~\ref{Fig:ipr0}(a). For strong disorder, $S_{\rm ent}$ saturates to a value of the order 1 with increasing the system size, indicating short-range entanglement in the MBL eigenstates, which is consistent with our basic assumption.

Next, we use the inverse participation ratio (IPR) as an intuitive, albeit somewhat indirect, test of the locality of operators  $\hat O_{{\cal L} i}$ from Eq.~(\ref{eq:eigenstates_connect}) which, when acting on products of eigenstates of systems $\cal{L},\cal{R}$, give eigenstates of $\cal LR$.
IPR for some state $|\Psi\rangle$ over a complete basis $|\alpha_i\rangle$ is defined as $\text{IPR}(|\Psi\rangle) = (\sum p_i^2)^{-1}$, where $p_i  = |\langle \Psi | \alpha_i\rangle|^2$ represents the probability of finding a state $| \alpha_i\rangle$. 
Defined in such a way, IPR takes values between $1$ and the Hilbert space dimension, and effectively tells us how many components have nonzero weight in the decomposition of the given state over the chosen complete basis. Fig.~\ref{Fig:ipr0}(b) shows the average IPR for the product $|\alpha\rangle \otimes |\beta\rangle$ of two random eigenstates of $\cal L$ and $\cal R$ subsystems  over the eigenstates $|\lambda\rangle$ of $\cal LR$. Value of IPR at strong disorder is very close to 1, indicating that the product of eigenstates of $\cal L$ and $\cal R$ is ``close" to the eigenstate of the full system $\cal LR$. Furthermore, IPR does not grow with $L$ for strong disorder, suggesting that the product of eigenstates of $\cal L$ and $\cal R$ differs from the eigenstate of the full system only near the boundary. 
 
To provide further support for our construction of the integrals of motion, we numerically implemented the projector operator similar to the one defined in Eq.~(\ref{eq:integral}). Every eigenstate $|\lambda \rangle$ of $\cal LR$ is labeled by its ``ancestor'' in $\cal L$ as in Eq.~(\ref{eq:eigenstates_connect}).   To find the  ancestor, we calculate the density matrix $\hat \rho_\lambda$ for the $\cal L$ subsystem from $|\lambda \rangle$. Using $\hat \rho_\lambda$,  we extract the probabilities of all eigenstates of $\cal L$ as $p_\alpha  = |\langle  \alpha|\hat \rho_\lambda | \alpha \rangle|^2$. In the limit of very strong disorder the typical value of the largest $p_\alpha$ is close to one~\cite{SOM}. Thus, the ``ancestor'' for $|\lambda\rangle$ is defined to be an eigenstate of $\cal L$ with the largest probability $p_\alpha$.  

\begin{figure}[t]
\begin{center}
\includegraphics[width=\linewidth]{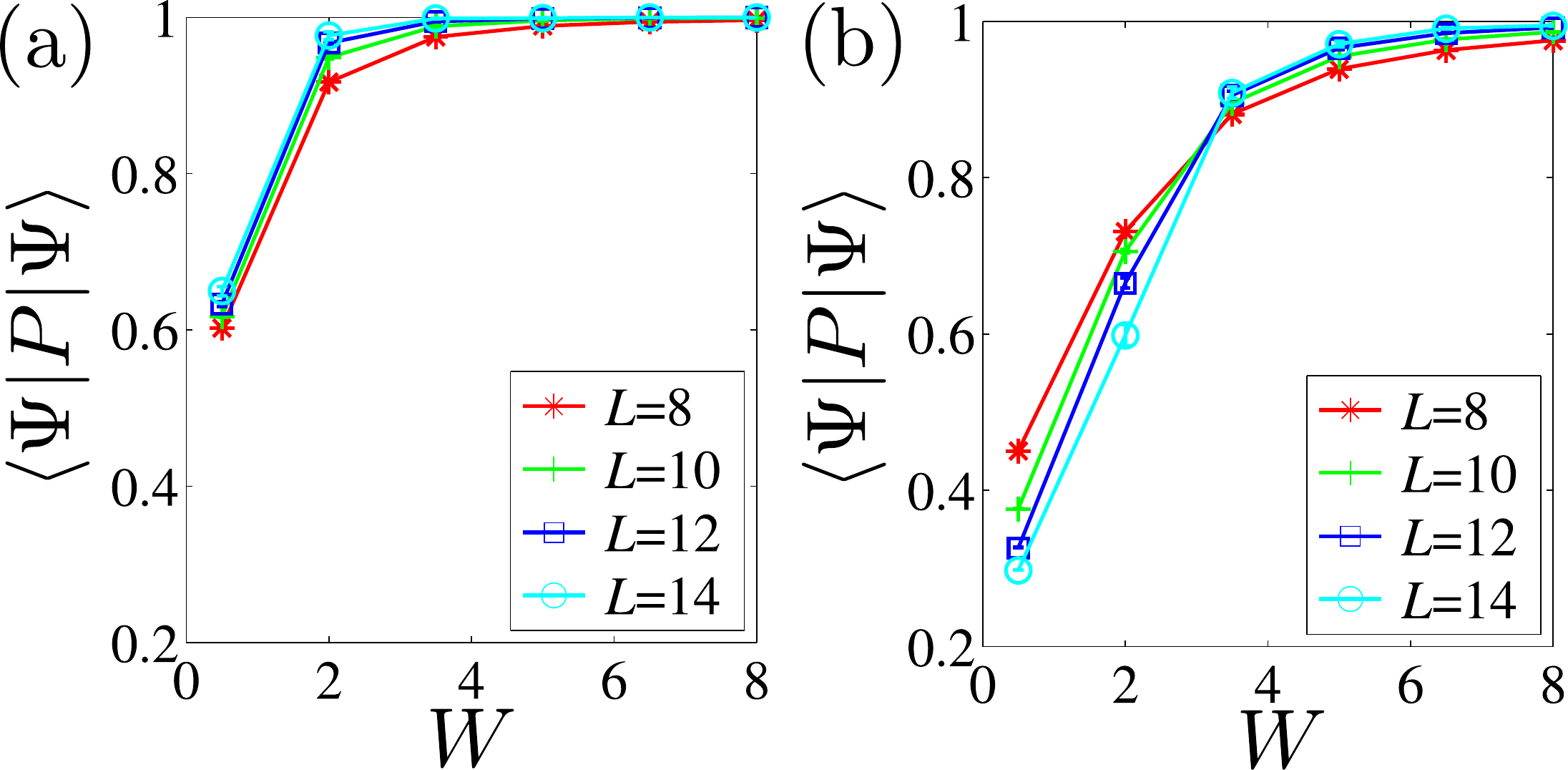}
\caption{ \label{Fig:ipr}  The weight of the perturbed eigenstate $|\lambda^\alpha\rangle$ in the subspace with index $\alpha$. For strong disorder, the action of the projector is contained within the subspace $\alpha$, irrespective of the interaction: case of no interaction, $V=0$ is shown in panel (a), and $V=1$ in panel (b). The weight increases with system size. For weak disorder ($W<3$), the presence of interactions causes the weight to decrease with the system size, suggesting the onset of the delocalized phase.  
}
\end{center}
\end{figure}

Although we do not assign labels for the right subsystem, such labelling is sufficient to implement the operator $\hat P_{\alpha} =  \sum_\beta \hat P_{\alpha\beta}$ as a projector onto the subspace of all eigenstates with the same label $\alpha$ for the $\cal L$. As a simple test, we study the locality of the projector~$\hat P_\alpha$: by construction it must have trivial action in the right subsystem. To test this property, we perturb some eigenstate with label $\alpha$, $|\lambda^\alpha\rangle$, at the right boundary, $|\psi^\alpha \rangle = (1/2+2\vec S_L\cdot \vec S_{L-1})|\lambda^\alpha\rangle$. Because we are interested in the weight of $|\psi^\alpha \rangle $ in the subspace with the same label $\alpha$, we plot the averaged $\langle \psi^\alpha| \hat P_\alpha |\psi^\alpha \rangle$ as a function of disorder in Fig.~\ref{Fig:ipr}. For strong disorder, even when the interaction strength is $V=1$, the perturbed state $|\psi^\alpha \rangle$ has almost all of its weight in the subspace with index $\alpha$, indicating that the degrees of freedom in the subsystem $\cal L$ are not affected by the perturbation acting on the subsystem $\cal R$. It is evident from Fig.~\ref{Fig:ipr}(b) that the weight within the subspace $\alpha$ grows as a function of system size at $W>W_{*}$, and decreases at $W<W_*$, where $W_*\approx 3$. Thus, $W_*$ gives an estimate of the MBL transition location in agreement with Ref.~\cite{PalHuse}. We note that the construction described above allows for more explicit tests to be done, which will be presented in future work~\cite{wetobe}. Additional numerical verifications of our central assumption can be found in~\cite{SOM}.

{\bf Discussion.} We established that the MBL phase is characterized by a number of local integrals of motion, supporting the hypothesis put forward in Ref.~\cite{Altman13}. This implies that the MBL phase does not thermalize, and only partial thermalization of initial product states, constrained by the local conservation laws, is possible. 

It should be noted that there are many ways to define local integrals of motion. For example, in certain problems~\cite{Huse13} it might be helpful to label the integrals of motion by a set of 1/2-pseudospins. Then, ${\cal M}=2^K$ possible values of a given integral of motion $\hat I_i$ can be viewed as states of $K$ pseudospins $\sigma_i^\eta$, $\eta=1,...K$. The $z$-projections of these pseudospins form a complete set of integrals of motion, and the Hamiltonian only involves $\sigma_{iz}^\eta$ operators and their products. Operators $\sigma_i^\eta$ can be viewed as effective degrees of freedom, in terms of which the dynamics becomes trivial: up-down states of spins are eigenstates, so time evolution can only lead to the dephasing between them. 

Another implication of our work concerns the structure of the MBL eigenstates: they are short-range entangled, obey the area law, and can be generally represented as a product of eigenstates of the subsystems of size $\gg \xi$ which have been locally ``corrected" near the boundaries with neighboring subsystems. This suggests an efficient numerical procedure for describing the MBL eigenstates in terms of matrix-product states. Starting from the product of eigenstates of decoupled blocks of size $\gg \xi$, entanglement between the blocks is introduced by the repeated action of the boundary terms in the Hamiltonian. The boundary terms generate only a finite-dimensional space, thus diagonalizing the boundary Hamiltonian for each finite-dimensional subspace, it should be possible to find the eigenstates of two coupled blocks, etc. 

Finally, our picture suggests a realistic route to extending coherence times in nearly isolated quantum systems, where decoherence is induced by interactions. Examples of such systems, in addition to systems of ultracold atoms, include nuclear spins and NV centers in diamond~\cite{Lukin06}. Assuming that one could induce strong static disorder leading to the many-body localization, the coherence time of a subsystem can be made very long. To achieve this, one needs to prepare a subsystem of size $\gg\xi$ (e.g., subsystem 1 in the above example), as well as its immediate neighborhood (e.g., subsystem 2) in some eigenstate. Then, local operations on the subsystem's degrees of freedom would couple states with different integrals of motion $I_1$, but with fixed values of $I_2$. Therefore, even though the rest of the system is in some complicated superposition state, it will only give rise to an exponentially weak dephasing, with the rate proportional to $\exp(-l/\xi)$. 

{\bf Acknowledgements.} We thank J. Moore for useful discussions. Research at Perimeter Institute is supported by the Government of Canada through Industry Canada and by the Province of Ontario through the Ministry of Economic Development \& Innovation. Z.P. was supported by DOE Grant No. DE-SC0002140. M.S. was supported by the National Science Foundation under Grant No. DMR-1104498. The simulations presented in this article were performed on computational resources supported by the High Performance Computing Center (PICSciE) at Princeton University.

{\it Note added.} During the completion of this manuscript, we became aware of a related work~\cite{Huse13} discussing the existence of local integrals of motion in the MBL phase.


\newpage 

\setcounter{figure}{0}
\makeatletter 
\renewcommand{\thefigure}{S\@arabic\c@figure}

\section*{Supplemental Online Material for ``Local Conservation Laws and the Structure of the Many-Body Localized States"}

Below, we present a number of numerical tests which support the central assumption that local perturbations lead only to local modifications of the eigenstates in the many-body localized phase. In addition, we show additional evidence for the viability of the numerical construction of the projector operator presented in the main text.

\subsection{Fidelity}

\begin{figure}[b]
\begin{center}
\includegraphics[width=\linewidth]{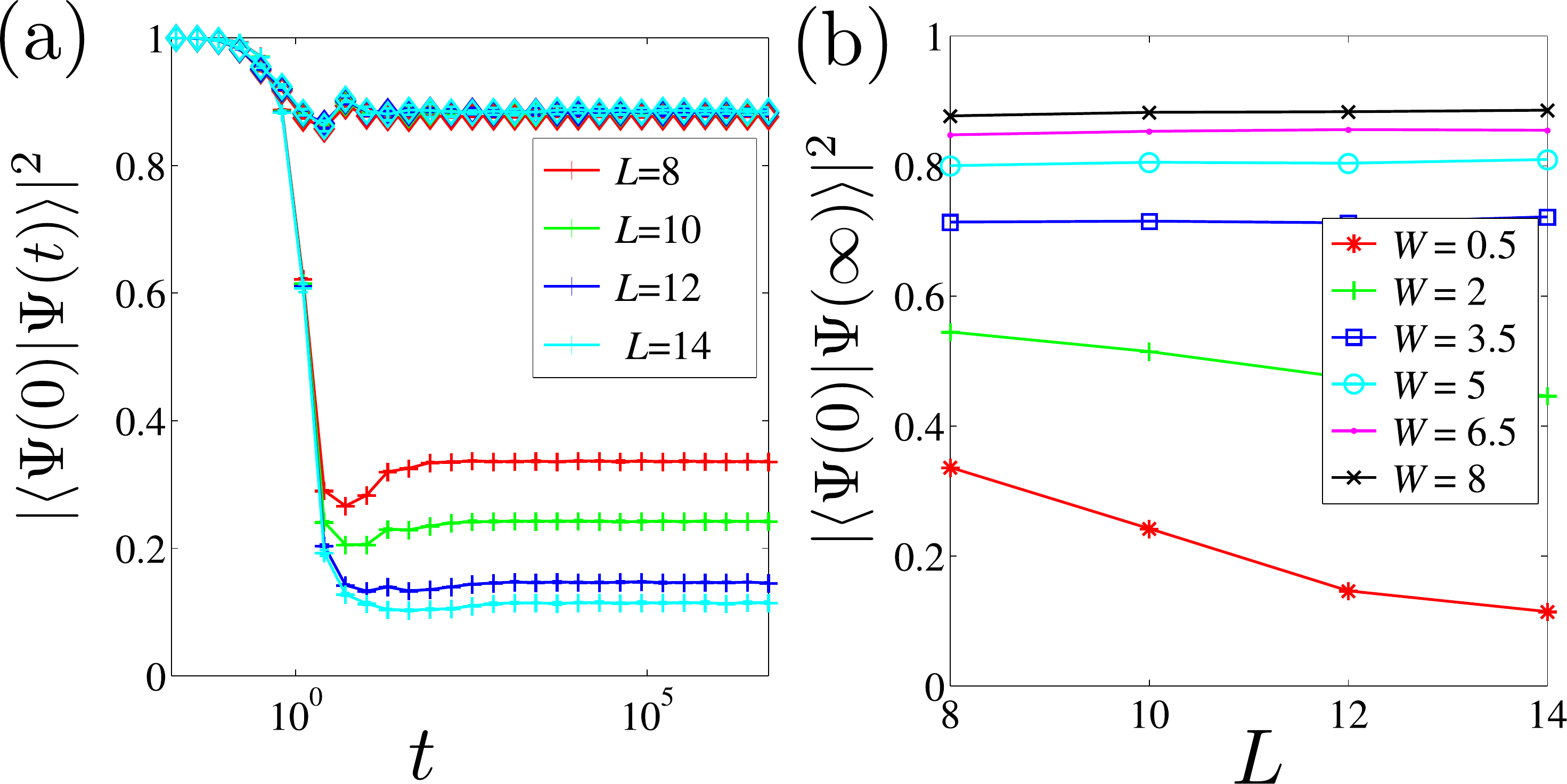}
\caption{Fidelity of an eigenstate perturbed at the right boundary. (a) Averaged fidelity as a function of time saturates to a value close to 1 for strong disorder and independent of system size (top curves, $W=8$). For weak disorder ($W=0.5$), fidelity saturates to a system-dependent value that decreases with $L$. (b) Saturated value of the fidelity displays a  crossover to the regime where fidelity is independent of system size for sufficiently strong disorder. Interaction strength is $V = 1$.}
 \label{Fig:fid} 
\end{center}
\end{figure}

First, we study fidelity, defined as the squared overlap of a given initial state with itself after time $t$, $F(t)=|\langle \psi(t)|\psi(0)\rangle|^2 \equiv |\langle \psi|e^{-iHt}|\psi\rangle|^2$. Initial state is chosen  to be an eigenstate $|\lambda_i\rangle$, perturbed at the right boundary $|\psi\rangle = (1/2+2\vec S_L\cdot \vec S_{L-1}) |\lambda_i\rangle$, where $\vec S_{L}$,$\vec S_{L-1}$ are the spin operators at the two rightmost sites. Fig.~\ref{Fig:fid}(a) shows the fidelity as a function of time, averaged over different initial states and realizations of disorder. For strong disorder, the saturated fidelity $F(\infty)$, Fig.~\ref{Fig:fid}(b), weakly depends on system size, demonstrating the local character of the perturbation introduced at the right boundary.

\subsection{Inverse participation ratios from the density matrix}

\begin{figure}[b]
\begin{center}
\includegraphics[width=\linewidth]{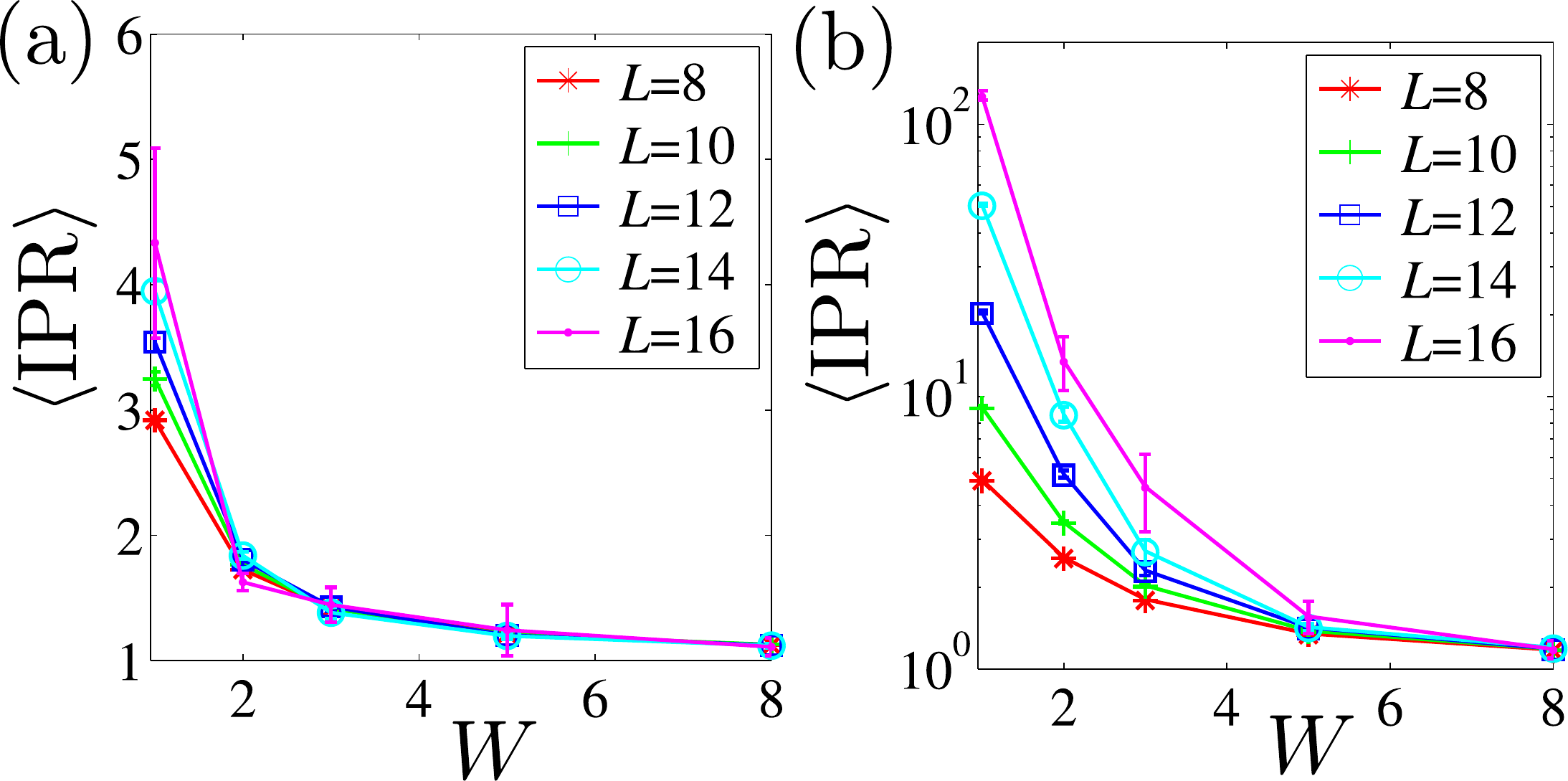}
\caption{Participation ratios for the eigenstate of a full system $\mathcal{L}\mathcal{R}$ over the eigenstates of left subsystem $\mathcal{L}$ remains close to one.   (a) For the non-interacting case IPR is always close to one.  (b) For the interacting system with $V = 1$ IPR also tends to one for sufficiently strong disorder. For weaker disorder, there is a crossover to a regime with longer range entanglement.\\ \ \\}
\label{Fig:iprb0}  
\end{center}
\end{figure}

In the main text, we have defined the inverse participation ratio (IPR) for some state $|\Psi\rangle$ over a complete basis $|\alpha_i\rangle$ as $\text{IPR}(|\Psi\rangle) = (\sum p_i^2)^{-1}$, where $p_i  = |\langle \Psi | \alpha_i\rangle|^2$ is the probability of finding an eigenstate $| \alpha_i\rangle$ in the decomposition of $|\Psi\rangle$. IPR, defined in such a way, takes values between $1$ and the dimension of the Hilbert space, and effectively tells us how many components have nonzero weight in the decomposition of the given state over a chosen complete basis. We have shown that, in the strong-disorder limit, the average IPR for the product of two random eigenstates of the $\cal L$ and $\cal R$,  over the eigenstates of $\cal LR$, approaches 1. 

\begin{figure*}[htb]
  \begin{minipage}[l]{\linewidth}
    \includegraphics[scale=0.5]{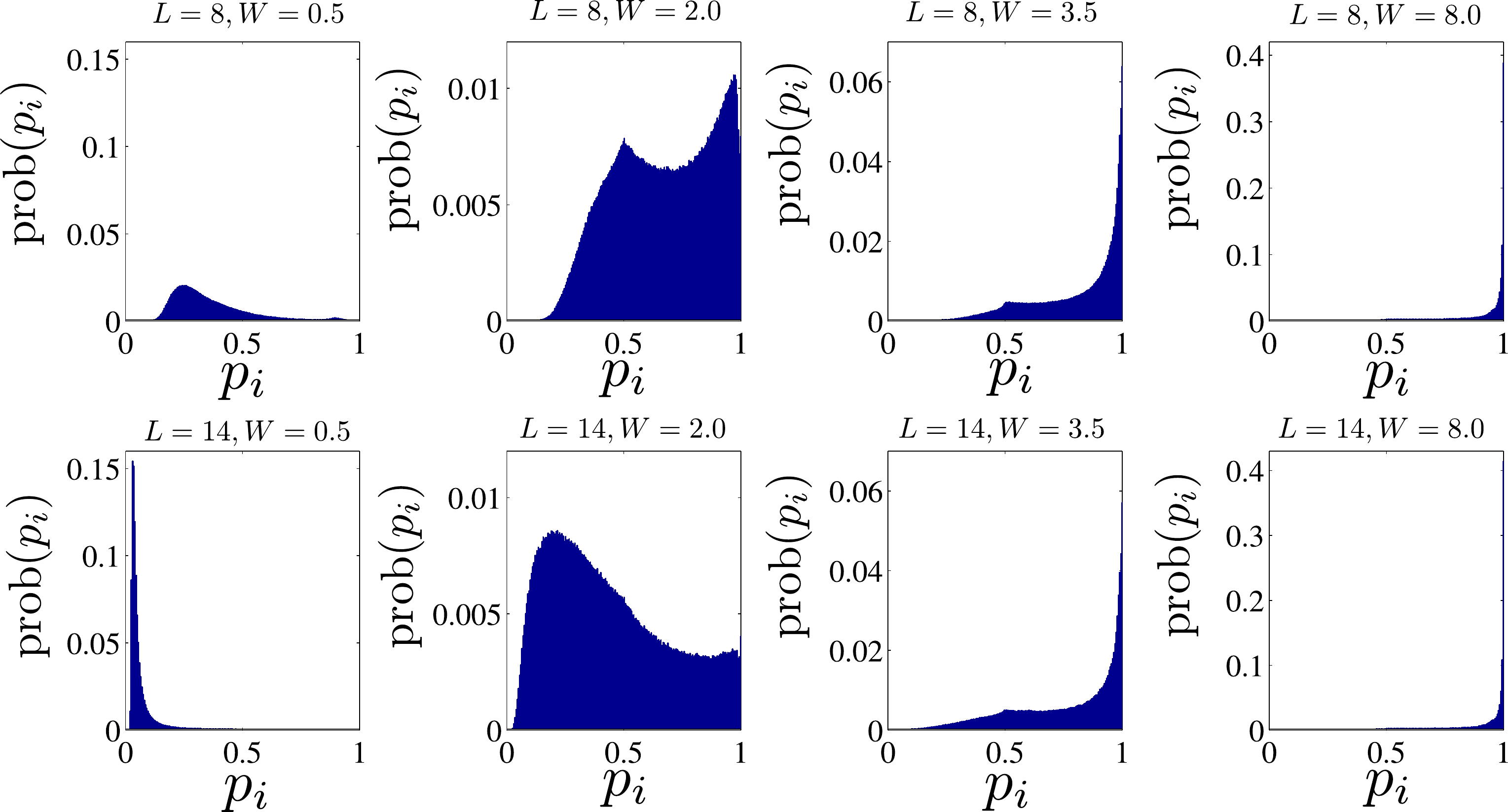}
  \end{minipage}
\caption{Histograms showing the distribution of the weights $p_i$. Top row is for $L=8$ spins, and the bottom is for $L=14$. The disorder strength,  $W$, is indicated above each plot. First two columns with $W=0.5$ and $W=2$ (corresponding to the delocalized phase) show that the median value of $p_\text{max}$ scales to zero in the thermodynamic limit. On the other hand, the last two columns ($W=3.5,8.0$) show that the $p_\text{max}$ is close to one for large disorder, thus confirming the ``similarity" of the given eigenstate of the full system to the product of the eigenstates of its subsystems. Interaction strength is $V = 1$ in all cases.}
\vspace{-0pt}
 \label{Fig:hist}
\end{figure*}
Additionally, here we show that IPR of the eigenstates of the full system over the eigenstates of $\cal L$ have similar behavior. Starting from a given eigenstate of $\cal LR$, we calculate the density matrix for the $\cal L$ subsystem.  From the density matrix $\hat \rho$, we can extract the probabilities of all eigenstates of $\cal L$ by defining  $p_i  = |\langle  \alpha^{\cal L}_i|\hat \rho | \alpha^{\cal L}_i\rangle|^2$. For the non-interacting case, the IPR defined in such a way remains close to unity, Fig.~\ref{Fig:iprb0}(a). More interestingly, in the interacting case ($V=1$), the average IPR also approaches unity for sufficiently strong disorder, Fig.~\ref{Fig:iprb0}(b). This suggests there is a single $p_{i_{\rm max}}=p_\text{max}$ dominating in the sum $\sum p_i = 1$. 

To quantify the last point, in Fig.~\ref{Fig:hist} we study the distribution of the weights $\{ p_i\}, i=1,\ldots,2^{L/2}$, for different system sizes and disorder strengths. Interaction strength is fixed at $V=1$, like in Fig.~\ref{Fig:iprb0}. In the delocalized phase, corresponding to small disorder ($W=0.5, 2.0$), the median value of $p_\text{max}$ scales down to zero with the increase in system size. In contrast, for large disorder ($W=3.5,8.0$), $p_\text{max}$ is close to one, thus confirming the assumption that a  given eigenstate of the full system is ``close" to the product of the eigenstates of its subsystems, and justifying the labeling scheme used to numerically construct the projector in the main text.


\begin{thebibliography}{16}
\bibitem[]{footnote1}
Localization length depends on the energy density, and $\xi$ should be viewed as the largest localization length. One may envision models where  different excitations are characterized by different localization lengths. In this case as well, $\xi$ should be chosen as the largest localization length.
\end{thebibliography}
\end{document}